\documentclass[a4paper, amsfonts, amssymb, amsmath,  reprint, showkeys, nofootinbib, oneside, doublecolumn, 10pt]{revtex4-1}

\usepackage{amsthm}
\usepackage{mathtools}
\usepackage{physics}
\usepackage{xcolor}
\usepackage{graphicx}
\usepackage[left=20mm,right=20mm,top=35mm,columnsep=15pt]{geometry} 
\usepackage{adjustbox}
\usepackage{placeins}
\usepackage[T1]{fontenc}
\usepackage{lipsum}
\usepackage{csquotes}
\usepackage[english]{babel}
\usepackage[utf8]{inputenc}
\usepackage[pdftex, pdftitle={Article}, pdfauthor={Author}, colorlinks=true,allcolors=blue]{hyperref} 
\setcitestyle{authoryear,open={(},close={)}} 

\begin{document}
\title{A.I. go by many names: \\ towards a sociotechnical definition of artificial intelligence}

\author{\normalsize{Johannes Dahlke}}
    \email[Correspondence email address: ]{j.dahlke@utwente.nl}
    \affiliation{University of Twente, HBE/BMS-ETM, The Netherlands}
    \affiliation{ETH Zurich, D-MTEC, KOF Swiss Economic Institute, Switzerland}
    \affiliation{ISTARI.AI, Germany}

\date{\today} 

\begin{abstract}
Defining artificial intelligence (AI) is a persistent challenge, often muddied by technical ambiguity and varying interpretations. Commonly used definitions heavily emphasize technical properties of AI but neglect the human purpose of it. This essay makes a case for a sociotechnical definition of AI, which is essential for researchers who require clarity in their work. It explores two primary approaches to define AI: the rationalistic, which focuses on AI as systems that think and act rationally, and the humanistic, which frames AI in terms of its ability to emulate human intelligence. By reconciling these approaches and contrasting them with landmark definitions, the essay proposes a sociotechnical definition that includes the three central aspects of i) technical functions, ii) human purpose, and iii) dynamic expectations. 
\end{abstract}


\maketitle
\section*{Introduction} \label{sec:outline}



I frequently encounter the question of what I mean when using the term \emph{artificial intelligence} (AI) in my research on its diffusion and application in society. The infuriating truth of the matter is that defining AI can be just as easy as it can be difficult. It can be easy in the sense that there are clear technical specifications for given algorithmic solutions but it becomes difficult to abstract those to a higher order without blurring lines. It can be easy in the sense that people intuitively know that a machine behaves intelligently when they experience it, and difficult in the sense that this intuition may differ across time and contexts \citep{caluori2023hey}. 

So, let us assume that we would want to define AI in a way as to resolve this conundrum. Likely, we would start by thinking about technology itself and could consider defining it as the "knowledge of how to fulfill certain human purposes in a specifiable and reproducible way" \citep[p.66]{Brooks.1980}. Here, we find two requirements that our definition of AI would need to meet: It would need to comprise i) technical specifications and ii) social purpose; making it what some would call `sociotechnical' \citep{cooper1971sociotechnical, Brooks.1980} or others `sociomaterial' \citep{Barad.1998, Barad.2007} in nature. 

Some technologies allow to be clearly defined in terms of their technical specifications and the specific purpose they address. Take a traditional (incandescent) light bulb, for example. Its purpose is to illuminate populated space by emitting light.\footnote{Note the difference between the goal of emitting light and the purpose of illuminating populated space for humans.} It does so by sending an electrical current through a tungsten filament encased by a glass bulb filled with inert gases. We may find light bulbs coming in different sizes, shapes and colors. We may even find bulbs that have substituted the tungsten filament with a microchip (which we call LED). Yet, the purpose remains the same and the technical specifications remain comparable.

AI technology cannot be defined just as easily. While there is an obvious difference between analogue and digital technologies and a difference between more or less complex technologies, the struggle to clearly define the technology is chiefly rooted in two issues of scope. Firstly, it is rooted the wide scope of applications areas for AI technology constituting its general purpose character \citep{Brynjolfsson.2021}. Secondly, it is rooted in the wide scope of technical specifications of algorithmic solutions \citep{Domingos.2015}. This results in the equally wide but popular consensus-term `AI', which acts as an "umbrella term for a range of algorithm-based technologies" \citep[p.6]{Leslie.2020} serving different purposes. 

Some have suggested to drop using the term `AI' entirely \citep{lanier2023there} and refer to more specific algorithmic technologies instead. While this makes sense form the technical perspective of a computer scientist, language is a collective phenomenon and the popularity of this particular term renders this impetus esoteric. In fact, umbrella terms such as AI serve an important function as mediators between science and society and allow social sciences to coordinate the study of large-scale phenomena beyond local contexts \citep{rip2013umbrella}, such as the macroeconomic implications of diffusing AI technology. Rather than lamenting its use, we, thus, may want to focus on defining it more clearly to increase its potency as a mediator.

For approaching a definition that fulfills our two requirements, it is helpful to understand how AI has been defined previously. Two different approaches of defining AI technology are of relevance: the rationalistic and the humanistic one. Both of which will contribute different parts to our quest of finding defining specifications and purpose of AI technology. 
\section*{Rationalism} \label{sec:develop}
The rationalistic approach focuses on the defining trait of AI as being capable of thinking rationally and performing rational action \citep{Poole.1998, Winston.1992}. From that perspective, AI is defined as (a system of) intelligent agents that perceive their environment, take action to maximize the value of an outcome based on a goal function, and learn to reason based on data or experience \citep{Russel.2021}. Framing AI as intelligent agents also helps to understand that, in contrast to more static computer programs, AI is expected to act with some degree of autonomy, persist over time, and adapt to changes in its environment \citep{Russel.2021}.

Take the example of linear regressions, which are a common starting point in every machine-learning lecture. Programs allowing to conduct such regressions technically meet the criteria of perceiving environments (data inputs), maximizing a goal function (produce estimates of relations between variables while minimizing standard errors), and adapting to changes in the environment (improve model fit with increasing amounts of data). However, they have no authority over which parameters to select in the regression model or how to interact them, which is a vital capacity if the complexity of the environment to be processed increases.

This means that if we'd define the goal of this technology to be the production of an estimate given a predefined model of the world, linear regressions would meet the formal definition of what constitutes machine learning or even AI. If we stipulate the goal to be the development of its own model of the world to inform decision making, it would not because it would fail to derive novel concepts and to produce actionable output rather than a statistical summary of results. Here, the difference between a computer program performing linear regressions and, let's say, a neural network model lies between model fitting and concept learning---with the latter arguably being a crucial capability to correctly \emph{perceive} an environment and remain adaptive to changes in this environment. Contrary to simple computer programs that can be mainly understood as tools, AI may be understood as its own entity capable of reasoning and deciding with a degree of autonomy.

However, the main issue with the rationalistic approach remains the technical broadness of the definition creating confusion around which specific technologies meet it. For the above example, it easy to see how the lines blur when discussing other types of regressions such as logistic ones or more sophisticated techniques to include the process of variable selection (e.g., LASSO regressions), capturing some of the agentic or decisive properties described.

While it is hard to judge objectively if the rather broad conceptual properties outlined in the rationalistic approach are met by specific computational techniques, defining what is AI and what is not, in practice, often is a subjective endeavor.
\section*{Humanism} \label{sec:conclusions}
As much as intelligent agents may be created in silico, they are employed in operando. In the original proposal for the Artificial Intelligence Project at the Dartmouth conference \citep{McCarthy.1955}, Nathaniel Rochester describes the difference between shallow computer programs and AI in terms of his experiences and expectations in interacting with it (p.7): 

\begin{quote}
\emph{
"In writing a program for an automatic calculator, one ordinarily provides the machine with a set of rules to cover each contingency which may arise and confront the machine. One expects the machine to follow this set of rules slavishly and to exhibit no originality or common sense. Furthermore one is annoyed only at himself when the machine gets confused because the rules he has provided for the machine are slightly contradictory. Finally, in writing programs for machines, one sometimes must go at problems in a very laborious manner whereas, if the machine had just a little intuition or could make reasonable guesses, the solution of the problem could be quite direct. This paper describes a conjecture as to how to make a machine behave in a somewhat more sophisticated manner [...]"}
\end{quote}

We can trace the humanistic framing of the AI back to the very origins of the technology. The famous Turing test proposes an experiment that may determine a machine to be intelligent if it can mislead a human interrogator posing written questions into assuming to be engaging in conversation with a human \citep{Turing.1950}. It makes clear that human capabilities and behavior serves as a consistent reference point to make sense of the emerging technology. The humanistic approach also framed the birth of the term artificial intelligence at the Dartmouth conference in 1955 \citep[p.11]{McCarthy.1955}, where scholars broadly defined AI as a "machine [behaving] in ways that would be called intelligent if a human were so behaving." 

In the context of studying AI diffusion and its socio-economic implications, the humanistic approach to understanding AI is advantageous in the sense that it helps to reconcile the central struggle of defining a dynamic technology in a consistent way. In terms of dynamism, it highlights the fact that the perception of what constitutes (artificial) intelligence need not be objective or purely technical but can be connected to subjective expectations and relative comparisons. What constitute artificial intelligence thus becomes time- and context dependent and can be formulated as a function of expectations \citep{Hidalgo}. These expectations may evolve as the sociotechnical reality that humans interacting with AI technology are embedded in changes. At the same time, the humanistic approach offers a consistent frame of reference (human behavior) where precisely defining AI in technical terms may otherwise be a moving target due to the pace of technological progress.

Finally, let us acknowledge that "the basic function of technology is the expansion of the realm of practical human possibility" \citep[]{Hannay.1980}. This notion leads us away from the technical specificities of a given technology and points us back at its purpose, which can only be understood in terms of how it interacts with our human-centered world. Against the backdrop of the humanistic framing, the search for purpose prompts us to ask why intelligent behavior is desirable in the first place. The most concrete answer to this question is that intelligent action creates value, leading us to the realm of economics and management studies. This literature places the most immediate value of AI in increases of human productivity \citep{Autor.2024} through automating and/or augmenting human labor \citep{Raisch.2021, DellAcqua.2023}.\footnote{Note that different ways of automating or augmenting labor may lead to different levels of productivity \citep{Brynjolfsson.2023}.} This can be achieved by reducing the need for humans to do perform laborious tasks, thus freeing up capacities for tasks of higher cognitive levels or by enhancing the way that human labor is conducted in collaboration with the machine, making this co-production efficient or innovative.

\section*{State of discourse} \label{sec:discourse}
Table \ref{tab:def} contrasts the scientific conceptualizations of AI with workable definitions that have been proposed in the socio-political discourse in recent years. Notice that most of them exhibit a rationalistic understanding of AI. The most popular definition by the OECD is closely related to Mitchell's notion of machine learning. Importantly, it emphasizes the systemic properties of applied AI in interaction with its users and environment.

While the rationalistic consensus unifies the definitions across regions on a more theoretical level, it also exemplifies a pragmatic issue of this approach. The definitions diverge on which specific technologies meet the general definition also practically. This is due to the algorithmic diversity starting at the paradigmatic difference between symbolic and statistical AI. The former---sometimes called good-old-fashioned AI \citep{Buchanan2005}---comprises what \cite{Domingos.2015} describes as \emph{tribes} of AI researchers focused on logic-based (symbolists) and knowledge-based approaches (analogizers). A common application spawning from this stream of research are expert systems that provide recommendations by following stipulated inference rules connected to a curated base of human knowledge. AI tribes have also formed around the paradigm of statistical AI comprising algorithms that are based on machine learning and deep neural networks (connectionists), evolutionary search- and optimization algorithms such as genetic algorithms (evolutionaries), and using probabilistic modeling techniques such as Bayesian networks (Baysesians).

Given this diversity, contemporary definitions diverge in the provided examples of technologies falling under the umbrella term AI (such as machine learning) and clearly remain non-exhaustive. The proposed European AI Act \citep{EC.2021} proposes the definition of AI to be complemented with a specific (continuously updated) list of technologies while the U.S. AI Bill of Rights \citep{US.2022} makes an intuitive but broad distinction between AI as automated systems and passive computing infrastructure. The definition of the UK's Alan Turing Institute---possibly due to its intellectual heritage---seems to be one of the only common definitions of AI that acknowledges humanistic elements related to applied AI. 
\section*{A sociotechnical definition of AI} \label{sec:sociotech}
The rationalistic and humanistic approach to defining AI technology both offer their advantages and drawbacks and are, thus, best understood as complements. Where the rationalistic approach can be technically most precise, higher-order technical definitions undermine this precision by creating technological grayzones. While the rationalistic approach grants AI a goal, it is lacks a definition of the pragmatic purpose of the technology. Where the humanistic approach may be arbitrary in its subjective judgements of what can be expected of an AI, and how these judgements may differ between individuals and contexts, it does help navigating technical grayzones. Much in the sense of 'you know it when you see it', the humanistic approach defines AI as function of expectations and purpose. A definition cannot blind out either/or perspective on AI as "technology must be sociotechnical rather than technical and [...] must include the managerial and social supporting systems necessary to apply it [...]" \citep[p.65]{Brooks.1980}. Thus, we'd come to propose that a sociotechnical definition of AI must reconcile the rationalistic and humanisitc approach:

\begin{quote}
\emph{Artificial intelligence (AI) refers to digital machine-based systems that process inputs, infer concepts, and generate outputs for explicit or implicit objectives. It can be expected to perceive and influence its environment to varying degrees of autonomy. The purpose of fulfilling these objectives is to automate human labor and/or augment human intelligence.}
\end{quote}

While one may disagree with this exact sociotechnical definition of AI, it is clear that any such definition must address three central aspects:
\begin{enumerate}
    \item It must specify a broad technical frame of how AI systems function while delineating AI from more traditional information technologies through the inclusion of concept learning.
    \item It must stipulate a clear human purpose of AI, placing the technology in the realm of human-centered societies. 
    \item It must allow for the fluidity of expectations that users have regarding the autonomy that AI technology exercises while serving its purpose.
\end{enumerate}

\bibliography{main.bib}

\appendix*
\begin{table*} \centering 
   \caption{Conceptualizations of Artificial Intelligence} 
   \label{tab:def} 
   \footnotesize
\begin{tabular}{p{0.2\textwidth}p{0.6\textwidth}p{0.1\textwidth}}
\textbf{Source} & \textbf{Definition} & \textbf{Approach}\\ \hline
Scientific definitions & & \\
\hline
\cite{Mitchell.2007} & "A computer program is said to learn from experience E with respect to some class of tasks T and performance measure P, if its performance at tasks in T, as measured by P, improves with experience E." & rationalistic \\
\cite{Russel.2021} & "[...] all computer programs do something, but computer \emph{agents} are expected to do more: operate autonomously, perceive their environment, persists over a prolonged time period, adapt to change, and create and pursue goals. A rational agent is one that acts so as to achieve the best outcome or, when there is uncertainty, the best expected outcome." & rationalistic \\
\cite{McCarthy.1955} & "For the present purpose the artificial intelligence problem is taken to be that of making a machine behave in ways that would be called intelligent if a human were so behaving." & humanistic \\
\cite{Kurzweil.1990} & "The art of creating machines that perform functions that require intelligence when performed by people." & humanistic \\
\hline
\multicolumn{3}{l}{Socio-political definitions} \\
\hline
 OECD AI definition \cite{OECDleg2019} & "An AI system is a machine-based system that, for explicit or implicit objectives, infers, from the input it receives, how to generate outputs such as predictions, content, recommendations, or decisions that can influence physical or virtual environments. Different AI systems vary in their levels of autonomy and adaptiveness after deployment." & rationalistic \\
EU AI Act \cite{EC.2021} & "The notion of AI system should be clearly defined to ensure legal certainty, while providing the flexibility to accommodate future technological developments. The definition should be based on the key functional characteristics of the software, in particular the ability, for a given set of human-defined objectives, to generate outputs such as content, predictions, recommendations, or decisions which influence the environment with which the system interacts, be it in a physical or digital dimension. AI systems can be designed to operate with varying levels of autonomy and be used on a stand-alone basis or as a component of a product, irrespective of whether the system is physically integrated into the product (embedded) or serve the functionality of the product without being integrated therein (non-embedded). The definition of AI system should be complemented by a list of specific techniques and approaches used for its development, which should be kept up-to–date in the light of market and technological developments through the adoption of delegated acts by the Commission to amend that list." & rationalistic \\
US AI Bill of Rights \cite{US.2022} & "An 'automated system' is any system, software, or process that uses computation as whole or part of a system to determine outcomes, make or aid decisions, inform policy implementation, collect data or observations, or otherwise interact with individuals and/or communities. Automated systems include, but are not limited to, systems derived from machine learning, statistics, or other data processing or artificial intelligence techniques, and exclude passive computing infrastructure. 'Passive computing infrastructure' is any intermediary technology that does not influence or determine the outcome of decision, make or aid in decisions, inform policy implementation, or collect data or observations, including web hosting, domain registration, networking, caching, data storage, or cybersecurity. Throughout this framework, automated systems that are considered in scope are only those that have the potential to meaningfully impact individuals’ or communities’ rights, opportunities, or access." & rationalistic \\
Canada's Artificial Intelligence and Data Act \cite{CA.2022} & “AI system” means a technological system that, autonomously or partly autonomously, processes data related to human activities through the use of a genetic algorithm, a neural network, machine learning, or another technique in order to: (a) generate content; or (b) make decisions, recommendations or predictions. & rationalistic \\
UK Information Commissioner's Office / Alan Turing Institute \cite{Leslie.2020} & "Artificial Intelligence (AI) can be defined in many ways. However, within this guidance, we define it as an umbrella term for a range of algorithm-based technologies that solve complex tasks by carrying out functions that previously required human thinking. Decisions made using AI are either fully automated, or with a ‘human in the loop’." & humanistic \\
\hline 
\end{tabular}
\end{table*}

\end{document}